\documentclass[11pt]{article}
\usepackage{geometry}                
\usepackage[small]{caption}
\usepackage{subfig}
\usepackage{graphicx}
\usepackage{amsmath}
\usepackage{amsfonts}
\usepackage{amssymb}
\usepackage{epstopdf}
\usepackage{mathrsfs}
\usepackage[T1]{fontenc}
\usepackage{feynmf}
\title{Construction of asymptotic fields for a charged particle}
\author{O.W. Greenberg \thanks{e-mail address, owgreen@umd.edu } \\
Steve Cowen \thanks{e-mail address, scowen@umd.edu} \\
\emph{Center for Fundamental Physics}\\
\emph{University of Maryland} \\
\emph{College Park, MD 20742, US} \\
and \emph{Helsinki Institute of Physics} \\
\emph{University of Helsinki} \\
\emph{Helsinki, FI-00014, Finland}\\
University of Maryland Preprint PP-012-011 }
\date{July 24, 2012}                                           

\begin{document}

\maketitle

\begin{abstract} 
Asymptotic fields do not exist in theories with massless particles and fields, because the vacuum matrix elements of products of the interacting fields in such theories do not have 
delta function or principal value singularities in momentum space. We remedy this problem by 
constructing a field for the charged particle that does have the required singularities in momentum space. We illustrate this construction in quantum electrodynamics (QED). 
\end{abstract}

\section{Introduction}
There are open questions about the description of charged particles in theories with long range 
interactions connected with massless fields and particles. For example, the asymptotic (in and out) states and fields of QED do not exist because of the long range Coulomb interaction. In this paper we construct a clothed field for a charged particle in QED. We show that the asymptotic limits exist for this clothed field. 

In a pioneering paper, Kulish and Faddeev~\cite{Kulish} showed that terms in the interaction of QED that have an annihilation and a creation operator of a charged particle give nonvanishing contributions in the limits $t \rightarrow \pm \infty$. These contributions come from the coupling of soft photons to the charged particles, and produce a branch cut at the (renormalized) mass of the charged field in the K\"all\'en-Lehmann (KL) weight of the two-point function, instead of a delta function. The analogous problem in the Feynman propagator is that there is a branch cut instead of a pole at the mass of the charged fields. Because of this branch cut, the asymptotic limits that would define the in and out fields for the charged fields do not exist. In addition these branch cuts produce infrared (IR) divergences in scattering amplitudes.

Lavelle and McMullen~\cite{Lavelle1} gave a method to clothe charged fields to create a 
gauge-invariant charged field that can be used to define asymptotic fields.  The KL weight of their clothed field has a delta function singularity at the mass of the charged particle. They also showed that the scattering amplitudes constructed with their clothed fields are free of IR 
divergences.~\cite{Lavelle2} 

Dirac~\cite{Dirac} and Creutz~\cite{Creutz} constructed a composite gauge invariant field,
\begin{align} \label{eq:1}
\psi_f(x)\equiv \text{exp}(-ie\int d^4y f^\mu (x-y)A_\mu (y))\psi (x)
\end{align}
for the field of the charged particles.
Dirac noted the existence of several choices for $f^\mu (x-y)$ that fulfill the gauge invariance condition, but are unphysical.   He argued the correct choice of $f^\mu (x-y)$ is the one that  produces the correct electric field. 

We also note the work of Stefanovich \cite{Stefanovich}, who reformulated QED to eliminate ultraviolet (UV) infinities in the S-matrix and Hamiltonian, and discussed applications to bound states.    He used the approach of~\cite{green} to find a finite Hamiltonian that is equivalent, for scattering, to the usual QED Hamiltonian.   

In this paper, in contrast to~\cite{green}, which eliminated the entire trilinear term in their scalar model, we eliminate only the soft photon part of the terms that give nonvanishing contributions in the limits $t \rightarrow \pm \infty$. This is the minimal reformulation of the Hamiltonian that allows the asymptotic limits for the in and out fields to exist. The interaction terms with soft photons are the terms that produce a branch point at the mass of the charged particle rather than a delta function in the KL weight (or a pole in the Feynman propagator). We show that the two-point function of the clothed charged field has an isolated singularity at the mass of the charged particle.

Our calculation gives a two-body interaction between charged particles associated with soft photons of momentum less that a given value, which, for illustration, we chose to be $\alpha m_e$.

\section{Preliminaries}
We repeat the discussion of~\cite{Kulish} to show why the terms with a creation and an annihilation operator of the charged particle together with a soft photon operator (which we call the problematic terms) produce a branch cut in the KL weight. For example, if the charged field represents the electron, the square of the mass of an electron produced on the vacuum by 
$ b^{\dagger}(\mathbf{p}) a^{\dagger}_{\mu}(\mathbf{k})$ is 
\[
m_e^2 + 2(\sqrt{\mathbf{p}^2 + m_e^2}|\mathbf{k}|-\mathbf{p} \cdot \mathbf{k}) \rightarrow m_e^2,
\mathbf{k} \rightarrow 0.
\]
Thus this state does not have a discrete mass.
The two-point function of the charged field will have a branch point at the square of the mass of the charged particle instead of an isolated singularity. When we eliminate the soft photons with momentum $|\mathbf{k}| \leq \alpha m_e$ from
the electron field, the square of the mass of $ b^{\dagger}(\mathbf{p}) a^{\dagger}_{\mu}(\mathbf{k})$ is greater than $m_e^2+2(\sqrt{\mathbf{p}^2 + m_e^2}-|\mathbf{p}|)\alpha m_e$; thus there is a gap between the mass of the electron and the mass of the lowest electron-photon state. The quantitative size of the gap is not important; only the existence of a finite gap is necessary to provide a discrete mass for the charged particle.

We noted that only certain terms in the interaction Hamiltonian destroy the gap.  The problematic terms survive when we take the asymptotic limits
$t\rightarrow \pm \infty$ and prevent the existence of a gap above $m_e^2$ in the KL weight.  These are the terms we remove with our clothing operator. To do this,
we absorb the soft photons for $|\mathbf{k}|< \alpha m_e$ in the definition of the electron operator, removing the low-momentum part of the terms in the trilinear interaction that do not produce (or annihilate) pairs, but keeping the remaining part of the electron-photon interaction.  

We begin with the QED Hamiltonian, $H=H_0+H_I$.  The free Hamiltonian is just 
$H_0= H_{0 f}+ H_{0 ph}$ where
\begin{align}
H_{0f}&=\int \frac{d^3p}{(2\pi)^{3/2}} E_p (b^{s \dagger }_pb^s_p+d^{s \dagger}_p d_p) \\
H_{0ph}&=-\int \frac{d^3k}{(2\pi)^{3/2}} |\mathbf{k}| a^\dagger_{\mu k} a^\mu_k
\end{align}
where $E_p=\sqrt{\mathbf{p}^2+m_e^2}$ and the creation and annihilation operators have the usual commutation relations
\begin{align} \label{eq:4}
\{b_p^s, b^{r \dagger}_k\}&=(2\pi)^{3/2} \delta^3(\mathbf{p-k})\delta^{rs} \\ \label{eq:5}
[a_{\mu p}, a_{\nu k}^\dagger]&=-(2\pi)^{3/2}\delta^3(\mathbf{p-k})g_{\mu \nu},
\end{align}
and the interaction Hamiltonian is
\begin{align} \label{eq:6}
H_I(t)=-e \int d^3x A_\mu (x) \bar{\psi}(x) \gamma^\mu \psi(x).
\end{align}
The creation and annihilation operators depend on three-momenta and spin; the sum over repeated spin indices is implied.  We do not include renormalization counter terms because they are irrelevant for the issues of this paper.  We divide our Hamiltonian, 
\begin{align}
H=H_{0f}+H_{0ph_{soft}}+H_{0ph_{hard}}+H_{I_{soft}} +H_{I_{hard}}
\end{align}
where
\begin{align}
H_{0ph_{soft}} &=-\int_0^{\alpha m_e} \frac{d^3k}{(2\pi)^{3/2}} |\mathbf{k}| a^\dagger_{\mu k}a^\mu_k, \\
H_{0ph_{hard}} &=-\int_{\alpha m_e}^{\infty} \frac{d^3k}{(2\pi)^{3/2}} |\mathbf{k}| a^\dagger_{\mu k}a^\mu_k,
\end{align}
and 
similarly $H_{I_{soft}}$ and $H_{I_{hard}}$, written in momentum space, involve integrals over small and large photon momenta, respectively.  When the asymptotic limit is taken, only the soft photon part of the interaction Hamiltonian will survive.

The asymptotic limit of this Hamiltonian has been discussed in both \cite{Kulish} and \cite{Lavelle1} and we only give a brief summary here.  We expand the fields in the usual way
\begin{align} \label{eq:10}
\psi(x)&=\int \frac{d^3 p}{(2\pi)^3}\frac{1}{\sqrt{2 E_p}}\sum_s(b_p^s u^s( p)\exp(-i p \cdot x)
+d^{s \dagger}_p v^s(p) \exp(i p\cdot x))\\
A_\mu(x)&=\int \frac{d^3 k}{(2\pi)^3}\frac{1}{\sqrt{2 |\mathbf{k}|}}(a_{\mu k} \exp(-i p \cdot x)+a_{\mu k}^\dagger \exp(i p\cdot x)).
\end{align}
With these expansions, we find eight terms in \eqref{eq:6}, each with some time dependence of the form 
$\exp(i f(E) t)$, where $f(E)$ involves sums and differences of energies .  We wish to know which terms will be significant and which terms will vanish in the asymptotic limit.  In this limit, $t\rightarrow \pm \infty $, therefore $f(E)$ must go to $0$ or else the term will vanish.  Only terms where $f(E)=\pm (E_{p+k}-E_p\pm |\mathbf{k}|)$ can survive the limit because for $|\mathbf{k}|\approx 0$, $f(E)\approx 0$.  The resulting Hamiltonian is
\begin{align}
H_{Ias}(t)=-\frac{e}{(2\pi)^{3/2}}\int \frac{d^3k d^3p}{\sqrt{2 |\mathbf{k}|} E( p)}p^\mu
( \exp(-i\frac{p\cdot k}{E( p)}t) \rho^\dagger ( p,k) a_{\mu k}+\exp(i\frac{p\cdot k}{E( p)}t) 
\rho ( p,k) a^\dagger_{\mu k})
\end{align}
where 
\begin{align}
\rho(p,k)=\sum_s(b^{s\dagger}_{p-k} b^s_p-d^{s\dagger}_ {p-k} d^s_p)
\end{align}
and we have used the small-k limit to simplify the energy sums in the exponentials: $E_{p-k}-E_p+|\mathbf{k}|\approx \frac{p\cdot k}{E_p}$ and we dropped $k$ in $E(p-k)$ in the denominator.  We also set $\mathbf{k} =\mathbf{0}$ in the Dirac wave functions so that the wave functions do not appear in $\rho(p,k)$. We plan to clothe the whole Hamiltonian in a later paper; however here we focus on the asymptotic Hamiltonian,  the free fermion Hamiltonian and the soft part of the free photon Hamiltonian. 

\section{The clothing transformation}
Following~\cite{green}, we introduce clothed operators related by a unitary 
clothing operator generated by $S$, to the
bare operators in the interaction picture Hamiltonian. (We will call this unitary clothing operator $S$ for short.)
We define a clothed operator $\mathbf{O}$ as
\begin{align}
\mathbf{O}=\exp(i S) O \exp(-iS).
\end{align}
where $O$ is the bare operator. We denote clothed operators with bold type.  Because the clothing operator is
unitary, clothed operators obey the same commutation relations given in \eqref{eq:4} and \eqref{eq:5} as the bare ones.  The clothed Hamiltonian as a function of clothed operators is equal to the bare Hamiltonian as a function of unclothed operators~\cite{green},
\begin{align} \label{eq:15}
\mathbf{H(b,b^\dagger,d,d^\dagger,a,a^\dagger)}=H(b,b^\dagger,d,d^\dagger,a,a^\dagger).
\end{align}
Since the clothing operator commutes with itself, it can be written as a function of clothed or bare fields:
\begin{align}
S(\mathbf{b,b^\dagger,d,d^\dagger,a,a^\dagger})=S(b,b^\dagger,d,d^\dagger,a,a^\dagger).
\end{align}
We use 
\begin{align} \label{eq:17}
\mathbf{H}(\mathbf{b,b^\dagger,d,d^\dagger,a,a^\dagger})&=\exp(-iS) H(\mathbf{b,b^\dagger,d,d^\dagger,a,a^\dagger})\exp(i S) \notag \\
&= H(\mathbf{b^\dagger,b,\cdots})-i[S,H(\mathbf{b^\dagger,b,\cdots})]+\frac{(-i)^2}{2!} [S,[S,H(\mathbf{b^\dagger,b,\cdots})]]+\ldots
\end{align}
to find the clothed Hamiltonian, 
where the bold fields represent clothed fields, the bold $\mathbf{H}$ represents a clothed Hamiltonian, and $H(\mathbf{b^\dagger,b,\cdots})$ is the bare Hamiltonian as a function of the clothed fields.  This relation is easily verified by inserting $1=\exp(-iS) \exp(iS)$ on both sides of each bare operator in the Hamiltonian.  We assume S has an expansion in powers of $\alpha$ and \eqref{eq:17} is a valid perturbative expansion.

The soft photon part of the asymptotic trilinear term is what interferes with the definition of in and out fields. Therefore we define our clothing to cancel only the soft photon part of this term, and further, only 
to the order $\alpha$ that we are calculating.  All integrals over photon momenta below are taken only over the range \{$k =0, \alpha m_e$\}. 

To remove the soft photon part of the asymptotic trilinear interaction we require $S$ to satisfy
\begin{align} \label{eq:18}
-i[S,H_{0}( \mathbf{b^\dagger,b,\cdots})]=-H_{Ias}(\mathbf{b^\dagger,b,\cdots}).
\end{align}
From \eqref{eq:17}, this will create the needed cancellation.  We use the equation of motion for $S$ in the interaction picture, 
\begin{align}
i\frac{dS}{dt}=[S,H_0]=-iH_{Ias}(\mathbf{b^\dagger,b,\cdots}),
\end{align}
to derive an expression for $S$:
\begin{align}
S(t)&=- \int dt  H_{Ias}(\mathbf{b^\dagger,b,\cdots})  \notag \\
&= \frac{ie}{(2\pi)^{3/2}}\int \frac{d^3k d^3p}{\sqrt{2 |\mathbf{k}|}}\frac{p^\mu}{p\cdot k}
( \exp(-i\frac{p\cdot k}{E( p)}t) \rho^{\dagger} ( p,k) a_{\mu k}-\exp(i\frac{p\cdot k}{E( p)}t)
\rho( p,k) a^\dagger_{\mu k}).
\end{align}
This clothing operator is nearly identical to the soft component of the "distortion operator" of 
\cite{Lavelle1}.

We find the order in $\alpha$ of terms involving integrals over $k$ with an upper limit of $\alpha m_e$.  Due to this limit, the photon creation and annihilation operators carry a power of 
$\alpha^{-3/2}$  (See Appendix A).  The lowest order term of the free fermion Hamiltonian, which does not involve a $k$ integral, is $O(\alpha^0)$.  The soft part of the free photon Hamiltonian is 
$O(\alpha^1)$; the asymptotic interaction is $O(\alpha^{3/2})$.  Each power of $S$ contributes 
$O(\alpha^{1/2})$.  We compute our clothed Hamiltonian to $O(\alpha^2)$.

We also used \eqref{eq:15} to find the clothed Hamiltonian by writing the bare operators of the original (bare) Hamiltonian in terms of clothed fields.  The relation between the clothed and bare operators is \begin{align}
\mathbf{b}_ p^s&=exp(iS) b^s_p exp(-iS) \\
\mathbf{a}^\mu_k&= exp(iS) a^\mu_k exp(-iS).
\end{align}
To first order in $S$,
\begin{align}
\mathbf{b}^s_p&= b^s_ p -i[b^s_p,S] \\
\mathbf{a}^\mu_k&=a^\mu_k-i[a^\mu_ k,S]. 
\end{align}
After performing the commutators, we find
\begin{align}
\mathbf{b}^s_p&=b^s_p+ \frac{ e}{(2\pi)^{3/2}}\int \frac{ d^3k}{\sqrt{2 |\mathbf{k}|}}
(\frac{p^\mu}{p \cdot k} b^s_{p-k} a_{\mu k} - \frac{\bar{p}^\mu}{\bar{p} \cdot k} b^s_{p+k}
a^\dagger_{\mu k}) \\
\mathbf{a}_{\mu k}&= a_{\mu k} -\frac{e}{(2\pi)^{3/2} \sqrt{2 |\mathbf{k}|}} \int \frac{d^3 p}{p\cdot k}p_\mu (b^{ s\dagger}_{p-k} b^s_p-d^{s \dagger}_{p-k} d^s_p). 
\end{align}
where $\bar{p}\equiv(p^0,\mathbf{p+k})$.  Since the clothing operator $S$ commutes with itself, it is easy to invert the clothing operation to get the bare fields in terms of clothed fields:  
\begin{align}
b^s_p&=\mathbf{b}^s_p- \frac{ e}{(2\pi)^{3/2}}\int \frac{ d^3k}{\sqrt{2 |\mathbf{k}|}}
(\frac{p^\mu}{p \cdot k} \mathbf{b}^s_{p-k} \mathbf{a}_{\mu k} - \frac{\bar{p}^\mu}{\bar{p} \cdot k} \mathbf{b}^s_{p+k}\mathbf{a}^\dagger_{\mu k})\\
a_{\mu k}&= \mathbf{a}_{\mu k} +\frac{e}{(2\pi)^{3/2} \sqrt{2 |\mathbf{k}|}} \int \frac{d^3 p}{p\cdot k}p_\mu (\mathbf{b}^{s \dagger}_{p-k} \mathbf{b}^s_p-\mathbf{d}^{s \dagger}_{p-k} \mathbf{d}^s_p) 
\end{align}

\section{Calculation of the clothed Hamiltonian}
In this section we use \eqref{eq:17} to clothe the Hamiltonian to $O(\alpha^2)$.  We can use \eqref{eq:18} to simplify the procedure (proof of \eqref{eq:18} for our explicit operators is given in Appendix B).  Since we calculate to $O(\alpha^2)$, we insert \eqref{eq:18} into \eqref{eq:17} and combine terms to find
\begin{align} \label{eq:29}
\mathbf{H}(\mathbf{b,b^\dagger,d,d^\dagger,a,a^\dagger})&= H_{\mathbf{b^\dagger,b,\cdots}}
-i[S,H_{\mathbf{b^\dagger,b,\cdots}}]+\frac{(-i)^2}{2!} [S,[S,H_{\mathbf{b^\dagger,b,\cdots}}]] \notag \\
&= H_{0f}+H_{0ph}-i[S,H_{Ias}]+\frac{(-i)^2}{2!} [S,-iH_{Ias}]  \notag \\
&= H_{0f}+H_{0ph}-\frac{i}{2}[S,H_{Ias}],
\end{align}
where we have dropped the term involving two commutators of $S$ with $H_{Ias}$ because it is higher order in $\alpha$.  The Hamiltonians in \eqref{eq:29} are all functions of clothed fields.  By design, we no longer have a trilinear term to order $\alpha^2$.  All that is left to find the clothed Hamiltonian is to find $[S,H_{Ias}]$.  Since both $S$ and $H$ are trilinear expressions in operators and have two terms each, the calculation is cumbersome.  For this reason, we will leave out much of the details.  We first define
\begin{align}
 \frac{i}{2}[H_{Ias},S]\equiv H_{self}+H_{qu},
 \end{align}
where $H_{self}$ is the bilinear self energy term and $H_{qu}$ represents the quartic interaction terms.  Before computing the commutator, we write the $H_{Ias}$ and $S$ in a simpler way:
\begin{align}
H_{Ias}&=-\frac{e}{(2\pi)^{3/2}}\int \frac{d^3k d^3 p}{\sqrt{2 |\mathbf{k}|} E_p}p^\mu (A_\mu+A_\mu^\dagger) \\
S_{Ias}&=\frac{i e}{(2\pi)^{3/2}}\int \frac{d^3k' d^3 p'}{\sqrt{2 |\mathbf{k}|'}}\frac{{p'}^\nu}{{p'}\cdot k'} (A'_\nu-{A'}_\nu^\dagger)
\end{align}
where 
\begin{align}
A_\mu&=(\mathbf{b}^{s\dagger}_p \mathbf{b} ^s_{p-k}-
\mathbf{d}^{s\dagger}_ p \mathbf{d}^s_{p-k})\mathbf{a}_{\mu k} \\
A'_\nu&=(\mathbf{b}_{p'}^{s\dagger}\mathbf{b}_{p'-k'}^s-\mathbf{d}_{p'}^{s\dagger}\mathbf{d}_{p'-k'}^s)\mathbf{a}_{\nu k'}.
\end{align}
Now the operator part of the commutator is
\begin{align}
[A_\mu+A_\mu^\dagger, A'_\nu-{A'}_\nu^\dagger]&=[A_\mu,A'_\nu]-[A_\mu,{A'}^\dagger_\nu]+[A_\mu^\dagger,A'_\nu]-[A_\mu^\dagger, {A'}_\nu^\dagger] \notag \\
&=([A_\mu,A'_\nu]+[A_\mu,A'_\nu]^\dagger)-([A_\mu,{A'}^\dagger_\nu]+[A_\mu,{A'}^\dagger_\nu]^\dagger)
\end{align}
The only commutators we need to compute are $[A_\mu, A'_\nu]$ and $[A_\mu, {A'}^\dagger_\nu]$.  We take hermitian conjugates to find the other two terms.  The first commutator is
\begin{align}
[A_\mu, A'_\nu]&= (2\pi)^{3/2}[ \delta^3(\mathbf{p-k-p'})(\mathbf{b}_p^{s\dagger} \mathbf{b}^s_{p'-k'}+\mathbf{d}_p^{s\dagger} \mathbf{d}^s_{p'-k'}) \notag \\
&\;\;\;\;\;\;\;\;\;\;\;\;\;-\delta^3(\mathbf{p'-k'-p})(\mathbf{b}_{p'}^\dagger \mathbf{b}_{p-k}+\mathbf{d}_{p'}^\dagger \mathbf{d}_{p-k})]\mathbf{a}_{\mu k} \mathbf{a}_{\nu k'},
\end{align}
and the second commutator is
\begin{align}
[A_\mu, {A'}^\dagger_\nu]&=(2\pi)^{3/2}[-(2 \pi)^{3/2}(\mathbf{b}_p^\dagger \mathbf{b}_{p'}+\mathbf{d}_p^\dagger \mathbf{d}_{p'})\delta^3(\mathbf{p-p'}) \notag \\
& \;\;\;\;+\mathbf{b}^\dagger_p \mathbf{b}^\dagger_{p'-k'} \mathbf{b}_{p-k} \mathbf{b}_{p'}+\mathbf{d}^\dagger_p \mathbf{d}^\dagger_{p'-k'} \mathbf{d}_{p-k} \mathbf{d}_{p'} \notag \\
&\;\;\;\;+\mathbf{b}^\dagger_p \mathbf{b}_{p-k} \mathbf{d}^\dagger_{p'-k'} \mathbf{d}_{p'} +\mathbf{d}^\dagger_p \mathbf{d}_{p-k} \mathbf{b}^\dagger_{p'-k'} \mathbf{b}_{p'} ]\delta^3(\mathbf{k-k'})g_{\mu \nu} \notag \\
& \;\;\;\;+[(\mathbf{b}^\dagger_p \mathbf{b}_{p'}+\mathbf{d}^\dagger_p \mathbf{d}_{p'}) \delta^3(\mathbf{p-k-p'+k'}) \notag \\
&\;\;\;\;-(\mathbf{b}^\dagger_{p'-k'} \mathbf{b}_{p-k}+\mathbf{d}^\dagger_{p'-k'} \mathbf{d}_{p-k})\delta^3(\mathbf{p-p'})] \mathbf{a}^\dagger_{\nu k'} \mathbf{a}_{\mu k}.
\end{align}
Collecting terms bilinear in operators, we find the self-energy correction term
\begin{align*}
H_{self}=e^2 m_e^2 \int \frac{d^3p d^3 k}{2 |\mathbf{k}|} \frac{1 }{E_p p \cdot k} (\mathbf{b}^\dagger_p \mathbf{b}_p+\mathbf{d}^\dagger_p \mathbf{d}_p).
\end{align*}
We will ignore this term in the next section and assume our energy is renormalized. Note that our clothing does not induce any photon self-energy correction term.

Collecting the remaining terms, it is clear $H_{qu}$ will be a sum of terms involving $b^\dagger b^\dagger b b$,  $b^\dagger b aa$, $b^\dagger b a^\dagger a^\dagger$, $b^\dagger b a^\dagger a$, all preceding terms with $b\rightarrow d$, and  $b^\dagger b d^\dagger d$.  No terms involving the photon creation or annihilation operators survive to $O(\alpha^2)$.  The only dependence on these operators in the clothed Hamiltonian will be in the free photon part.  We have decoupled the soft photons.  The final quartic interaction Hamiltonian is
\begin{align} \label{eq:38}
H_{qu}&= -\frac{e^2}{2 (2\pi)^{3/2}}\int d^3 k d^3 p d^3 p' K(\mathbf{k},\mathbf{p},\mathbf{p'})
[(\mathbf{b}^{s\dagger}_p \mathbf{b}^{r\dagger}_{p'-k} \mathbf{b}^s_{p-k} \mathbf{b}^r_{p'}+(\mathbf{b} \rightarrow \mathbf{d})) \notag \\
&\;\;\;\;\;\;\;\;\;\;\;\;\;\;\;\;\;\;\;\;\;\;\;\;\;\;\;\;\;\;\;\;\;\;\;\;\;\;\;\;\;\;\;\;\;\;\;\;\;\;\;\;\;\;\;\;\;+(\mathbf{b}_p^{s \dagger}\mathbf{b}^s_{p-k}\mathbf{d}^{r \dagger}_{p'-k}\mathbf{d}^r_{p'} 
+ \mathbf{b} \leftrightarrow \mathbf{d})]
\end{align}
where
\begin{align}
K(\mathbf{k},\mathbf{p},\mathbf{p'})=\frac{p\cdot p' (E_{p'}p+E_p p')\cdot k}{2 |\mathbf{k}|E_p E_{p'} p \cdot k p' \cdot k }
\end{align}
This quartic interaction is the contribution to the asymptotic interaction, $H_{Ias}$, from soft photons with momentum less than $\alpha m_e$.

\section{Equation of motion}
With the clothed Hamiltonian, we can find the equations of motion for the clothed interacting creation and annihilation operators.  We start by defining
\begin{align}
\mathbf{b}^s(\mathbf{p},t)\equiv \mathbf{b}^s_{p}\exp(-i E t).
\end{align}
We will find the Heisenberg equation of motion for this momentum and time dependent field.  We begin by finding the commutator of $\mathbf{b}_p^s$ with the clothed Hamiltonian.  The commutator with the free Hamiltonian is trivial
\begin{align}
[\mathbf{b}_p^s,H_0]=E_p \mathbf{b}^s_p,
\end{align}
and the commutator with the first quartic term in \eqref{eq:38} is
\begin{align}
[\mathbf{b}^s_q,H_{qu}|_1] &=-\frac{e^2}{2(2\pi)^3}\int d^3 k d^3 p( K(\mathbf{k},\mathbf{q},\mathbf{p})
 \mathbf{b}^{r \dagger}_{p-k} \mathbf{b}^s_{q-k} \mathbf{b}^r_{p}-K(\mathbf{k},\mathbf{p},\mathbf{q+k})
 \mathbf{b}^{r \dagger}_p \mathbf{b}^r_{p-k} \mathbf{b}^s_{q+k}) \notag \\
&= \frac{e^2}{2(2\pi)^3}\int d^3 k d^3 p (K(\mathbf{k},\mathbf{q},\mathbf{p})+K(-\mathbf{k},\mathbf{p-k},\mathbf{q-k}))\mathbf{b}^{r \dagger}_{p-k} \mathbf{b}^r_{p} \mathbf{b}^s_{q-k}.
\end{align}
  The commutator with the second quartic term in the Hamiltonian vanishes trivially.  The commutator with the third and fourth quartic terms are
\begin{align}
[\mathbf{b}^s_q,H_{qu}|_{3,4}] &=-\frac{e^2}{2(2\pi)^3}\int d^3 k d^3 p (K(\mathbf{k},\mathbf{q},\mathbf{p}) \mathbf{b}^{s}_{q-k} \mathbf{d}^{r \dagger}_{p-k} \mathbf{d}^r_{p}+K(\mathbf{k},\mathbf{p},\mathbf{q+k})  \mathbf{b}^{s}_{q+k} \mathbf{d}^{r \dagger}_{p} \mathbf{d}^r_{p-k}) \notag \\
&= -\frac{e^2}{2(2\pi)^3}\int d^3 k d^3 p (K(\mathbf{k},\mathbf{q},\mathbf{p})+K(-\mathbf{k},\mathbf{p-k},\mathbf{q-k}) ) \mathbf{b}^{s}_{q-k} \mathbf{d}^{r \dagger}_{p-k} \mathbf{d}^r_{p}.
\end{align}
Using the above commutators, our equation of motion is
\begin{align} \label{eq:44}
i \frac{d \mathbf{b}^s(\mathbf{q},t)}{dt}&=  E_q \mathbf{b}^s(\mathbf{q},t)+ \frac{e^2}{2 (2\pi)^3} \int d^3 k d^3 p \; T (\mathbf{k,q,p})\boldsymbol{\rho}(\mathbf{p,k},t) \mathbf{b}^{s }(\mathbf{q-k},t)
\end{align}
where
\begin{align}
T (\mathbf{k,q,p})=K(\mathbf{k},\mathbf{q},\mathbf{p})+K(-\mathbf{k},\mathbf{p-k},\mathbf{q-k}).
\end{align}
Since the clothed Hamiltonian is symmetric under $b\leftrightarrow d$ interchange, we find the equation of motion for $d$ by interchanging $b$ and $d$ in \eqref{eq:44},
\begin{align} \label{eq:46}
i \frac{d \mathbf{d}^s(\mathbf{q},t)}{dt}&=  E_q \mathbf{d}^s(\mathbf{q},t)- \frac{e^2}{2 (2\pi)^3} \int d^3 k d^3 p \; T (\mathbf{k,q,p})\boldsymbol{\rho}(\mathbf{p,k},t) \mathbf{d}^{s }(\mathbf{q-k},t)
\end{align}

We now find a perturbative solution to these equations in powers of $e^2$.  We begin by expanding the time dependent operator
\begin{align} \label{eq:47}
\mathbf{b}^{s}(\mathbf{q},t)&=\mathbf{b}^{s \;in}(\mathbf{q},t)+e^2 \mathbf{b}^{(2)s}(\mathbf{q},t)+e^4 \mathbf{b}^{(4)s}(\mathbf{q},t) +\ldots \\ \label{eq:48}
\mathbf{d}^{s}(\mathbf{q},t)&=\mathbf{d}^{s \;in}(\mathbf{q},t)+e^2 \mathbf{d}^{(2)s}(\mathbf{q},t)+e^4 \mathbf{d}^{(4)s}(\mathbf{q},t) +\ldots
\end{align}
The first term in the expansion is the annihilation operator for an asymptotic in field, which obeys the free equation of motion.  The time dependence of the higher order terms is unknown at this point, but we can assign them time dependence of the form $\exp(-i E^{(n)} t)$, where $E^{(n)}$ is the energy associated with the nth order and will be determined during our perturbative approach, i.e.
\begin{align}
\mathbf{b}^{(n)s}(\mathbf{q},t)\equiv\mathbf{b}^{(n)s}(\mathbf{q})\exp(-i E^{(n)} t).
\end{align}
To solve \eqref{eq:44} and \eqref{eq:46}, we use \eqref{eq:47} and \eqref{eq:48} for $\mathbf{b}$ and $\mathbf{d}$ and collect terms of the same power in $e$.  Since we have already identified the first term with the in field annihilation operator, we know its time dependence is $\exp(-iE_q t)$, and therefore at $O(e^0)$, we have an identity,
\begin{align}
E_q \mathbf{b}^{s \; in}_{q}=E_q \mathbf{b}^{s \; in}_{q}.
\end{align}
At $O(e^2)$, we find
\begin{align} 
(E^{(2)}-E_q) \mathbf{b}^{(2)s}(\mathbf{q},t)&= \frac{e^2}{2 (2\pi)^3} \int d^3 k d^3 p \; T (\mathbf{k,q,p}) \boldsymbol{\rho}^{in}(\mathbf{p,k},t) \mathbf{b}^{s \; in}(\mathbf{q-k},t) \\ \notag \\ \label{eq:52}
(E^{(2)}-E_q) \mathbf{b}^{(2)}{}^s(\mathbf{q})&= \frac{e^2}{2 (2\pi)^3} \int d^3 k d^3 p \; T (\mathbf{k,q,p}) \boldsymbol{\rho}^{in}(\mathbf{p,k}) \mathbf{b}^{s \; in}_{\mathbf{q-k}}  \notag \\
&\;\;\;\;\;\;\;\;\;\;\;\;\;\;\;\;\; \times \exp (i (E^{(2)}-(E_p+E_{q-k}-E_{p-k}))t)
\end{align}
where
\begin{align}
\boldsymbol{\rho}^{in}(\mathbf{p,k},t) &\equiv \mathbf{b}^{s \, in \dagger}(\mathbf{p-k},t) \mathbf{b}^{s\, in}(\mathbf{p},t)-\mathbf{d}^{s\, in \dagger}(\mathbf{p-k},t) \mathbf{d}^{ in \,s}(\mathbf{p},t).
\end{align}
There is no time dependence on the left hand side of \eqref{eq:52}, so the time dependence on the right hand side must vanish, i.e. $E^{(2)}=E_p+E_{q-k}-E_{p-k}$. Using this in the equation gives
\begin{align} \label{eq:54}
\mathbf{b}^{(2)s}(\mathbf{q})&=\frac{e^2}{2 (2\pi)^3}  \int d^3 k d^3 p \frac{\; T (\mathbf{k,q,p}) \boldsymbol{\rho}^{in}(\mathbf{p,k}) \mathbf{b}^{s \; in}_{\mathbf{q-k}} }{E_p+E_{q-k}-E_{p-k}-E_q } .
\end{align}
where we have again kept only the lowest order in $k$ to arrive at \eqref{eq:54}.  We find the equation for $d$ by taking $b \leftrightarrow d$.  Calculating to $O(e^4)$ is more laborious, and beyond the scope of this paper.  It requires computing a sum of terms, in each of which all the creation and annihilation operators are taken to be in fields except for one, which is taken to be the $O(e^2)$ term.  Since the $O(e^2)$ term has two terms itself, there are 12 terms before any contractions take place.  This calculation can be done, but it is not necessary for this work.

We have found an expansion of the clothed interacting fermion annihilation operator (and creation operator if we take the Hermitian conjugate of \eqref{eq:47}) in terms of normal ordered in field operators, which annihilate the vacuum and obey the free equation of motion.  This expansion does not involve any photon operators, nor will it at any order, because of the form of the clothed Hamiltonian.  

One thing that is worth studying is the clothed two-point function,
\begin{align}
&<\Omega|\{\bar{\boldsymbol{\psi}}_{\alpha}(x), \boldsymbol{\psi}_\beta(y) \} |\Omega> .
\end{align}
where 
\begin{align}
\boldsymbol{\psi}_\alpha(x)=\exp(iS) \psi(x) \exp (-iS).
\end{align}
The clothed field can be expanded in terms of the clothed operators and the expansion has the same form as \eqref{eq:10}.  We can then expand the clothed interacting operators in terms of clothed in field operators via \eqref{eq:47} and \eqref{eq:48}, inserting \eqref{eq:54} where necessary.  
From \eqref{eq:54}, we see that the second order contribution is expressed in terms of normal ordered in field creation and annihilation operators which annihilate the vacuum; therefore, they do not contribute to the two-point function.  Even if we include higher order terms, they will be normal ordered operators, and also will not contribute to the two-point function.  We still have a contribution from the leading order in field which gives the usual pole at the mass of the fermion squared, but all other terms vanish once we have absorbed the soft photons in the definition of the electron operator.  We still need to take the "hard" part of the Hamiltonian, in which a trilinear term still exists, into account.  This part of the Hamiltonian creates a branch cut in the spectral density beginning after some gap.  We can determine the size of this gap kinematically.  The invariant mass of the lowest order multi-particle state is
\begin{align}
q^2&=(E_p+|\mathbf{k}|)^2-(\mathbf{p+k})^2 \notag \\
&=m_e^2+2|\mathbf{k}|(E_p-\mathbf{p\cdot \hat{k}})
\end{align}
where $\hat{\mathbf{k}}$ is a unit vector in the direction of the photon momentum.  Our "hard" photons must have $|\mathbf{k}|>\alpha m_e$, so as long as $|\mathbf{p}|<\infty$, there exists a gap
\begin{align}
q^2-m_e^2=2|\mathbf{k}|(E_p-\mathbf{p\cdot \hat{k}})
\end{align}

We have suppressed the distribution aspects of our calculations. From the point of view of distribution theory, our formulas should be taken as distributions and should be integrated with a function from a Schwartz space~\cite{sch} such as $\cal{S}$. Then sets of measure zero or values in the
limit of momenta becoming infinitely large are not relevant. Thus it is not relevant that if
$p$ and $k$ are parallel, i.e. $\mathbf{\hat{p}\cdot \hat{k}}=1$, the gap above mass $m^2$ can vanish, since in the large $|\mathbf{p}|$ limit the gap goes as 
$m^2 |\mathbf{k}|/|\mathbf{p}|$.

\section{Conclusion and future work}
We reformulated the Hamiltonian of QED to eliminate the part of the asymptotic interaction, 
$H_{Ias}$, with soft photons that produces a branch cut at the mass of the electron. The clothed charged fields have absorbed the soft photons that prevented the proper definition of in and out fields. The reformulated charged fields have a gap between the mass shell single-particle singularity and the many-particle branch cut in the KL weight. The effects of soft photons that originally appeared in the trilinear interaction terms now appear in quadrilinear interaction terms. These quadrilinear interaction terms do not produce a branch cut in the KL weight, and we expect that they will not lead to infrared divergences in scattering amplitudes. 

We used our clothed Hamiltonian to find an equation of motion for the clothed creation and annihilation operators.  After finding these equations, we solved them perturbatively in powers of $e^2$.  From this process, we were able to find an expansion of the clothed interacting operators in terms of in field operators that annihilate the vacuum.  The $O(e^2)$ terms of the expansion contain normal ordered in field creation and annihilation operators with at least one annihilation and one creation operator.  Higher order terms will also be normal ordered.  Due to this normal ordering, these higher order terms did not contribute to the clothed two-point Green's function.  The combination of the normal ordering and the lack of photon operators in the expansions shown in \eqref{eq:47} and \eqref{eq:48} produces a gap between the mass shell singularity at $m_e^2$ and the beginning of the multi-particle branch cut in the KL weight.  From our calculations we have found the lowest-order contribution to the two-particle interaction from soft photons.

We plan to use our (possibly modified) clothing technique to investigate the effects of clothing the constituents of bound states.  

\section*{Appendix A: $\alpha$ power counting}
We have integrals of the form
\begin{align*}
\int_0^{\alpha m} d^3 k F(\mathbf{k}) a_\mu (\mathbf{k}).
\end{align*}
We need a consistent way to keep track of the order in $\alpha$ of such an expression.  Since 
$k$ is restricted to the range $[0,\alpha m]$, we expect that this integral is $O(\alpha^3)$ times the power of $\alpha$ in $F(\mathbf{k})$, if we neglect 
$a_\mu (\mathbf{k})$ in our power counting.  If we take a commutator between our integral and 
$a^\dagger_\nu (\mathbf{p})$, we find
\begin{align*}
[\int_0^{\alpha m} d^3 k F(\mathbf{k}) a_\mu( \mathbf{k}),a^\dagger_\nu (\mathbf{p})]=-g_{\mu \nu} F(\mathbf{p})
\end{align*}
where $\mathbf{p}$ is also restricted to the range $[0,\alpha m]$.  This term seems to be of the same order 
in $\alpha$ as $F( \mathbf{p})$.  Thus, if we ignore the $\alpha m$ dependence of $a_\mu (\mathbf{k})$, we have lost a factor of $\alpha^3$.
 
To take account of the $\alpha m$ dependence of $a_\mu (\mathbf{k})$, we change the range of $\mathbf{k}$ from 
$[0,\alpha m]$ to $[0,1]$ in a new variable $\hat{\mathbf{k}}= (\alpha m)^{-1} \mathbf{k} $.  Then the integral becomes
\begin{align*}
(\alpha m)^3 \int_0^1 d^3 \hat{\mathbf{k}} F(\alpha m \hat{\mathbf{k}}) a_\mu(\alpha m \hat{\mathbf{k}})
\end{align*}
The commutation relations for operators with such arguments are
\begin{align*}
[a_\mu(\alpha m \hat{\mathbf{k}}),a^\dagger_\nu (\alpha m \hat{\mathbf{p}})]&=-g_{\mu \nu}\delta (\alpha m (\hat{\mathbf{k}}-\hat{\mathbf{p}})) \\
[(\alpha m)^{3/2} a_\mu(\alpha m \hat{\mathbf{k}}), (\alpha m)^{3/2}a^\dagger_\nu (\alpha m \hat{\mathbf{p}})]&=-g_{\mu \nu}\delta^3 ( \hat{\mathbf{k}}-\hat{\mathbf{p}}) \\
[\hat{a}_\mu( \hat{\mathbf{k}}), \hat{a}^\dagger_\nu ( \hat{\mathbf{p}})]&=-g_{\mu \nu}\delta^3 ( \hat{\mathbf{k}}-\hat{\mathbf{p}}) 
\end{align*}
where we have defined $\hat{a}_\mu(\hat{\mathbf{k}})=(\alpha m)^{3/2} a_\mu(\mathbf{k})$, and we see that $\hat{a}_\mu$ satisfies the usual commutation relations.  Now our integral can be written as
\begin{align*}
(\alpha m)^{3/2} \int_0^1 d^3 \hat{k} F(\alpha m \hat{\mathbf{k}}) \hat{a}_\mu( \hat{\mathbf{k}})
\end{align*}
and it is clear that our integral is $O(\alpha^{3/2})$ times the power of $\alpha$ in $F(\alpha m \hat{\mathbf{k}})$.  Now if we commute with $a^\dagger_\nu( \mathbf{p})=(\alpha m)^{-3/2}\hat{a}^\dagger_\nu( \hat{\mathbf{p}})$, we find
 \begin{align*}
[(\alpha m)^{3/2}\int_0^{1} d^3 \hat{k} F(\alpha m \hat{\mathbf{k}}) \hat{a}_\mu(\hat{\mathbf{k}}),(\alpha m )^{-3/2}\hat{a}^\dagger_\nu( \hat{\mathbf{p}})]=-g_{\mu \nu} F( \alpha m \mathbf{p})
\end{align*}
and the power counting is consistent.  Each $a_\mu (\mathbf{k})$ and $a^\dagger_\mu (\mathbf{k})$ effectively carries a power of $(\alpha m)^{-3/2}$.

A quicker way to see this is to note that if we take $\hbar=1$, the free Hamiltonian is
$H=\int d^3 k |\mathbf{k}| a^{\dagger}_\mu(\mathbf{k}) a^\mu(\mathbf{k})$; since the dimensions of $H$ and $|\mathbf{k}|$ are the same, $a$ and $a^{\dagger}$ must each carry dimensions $\mathbf{k}^{-3/2}$.

\section*{Appendix B: Checking (18) explicitly}
We check \eqref{eq:18} explicitly in here.  We focus on the $\mathbf{b}^\dagger \mathbf{b}$ part of the free fermion Hamiltonian because the $\mathbf{d}^\dagger \mathbf{d}$ part is clothed in the same way.  The commutator of $S$ with $H_{0f}|_{b}$, 
where $H_{0f}|_{b}$ is the $\mathbf{b}^\dagger \mathbf{b}$ term, is
\begin{align}
[S,H_{0f}|_{b}] &=-\frac{ie}{(2\pi)^{3/2}} \int \frac{ d^3 k d^3 p}{\sqrt{2 |\mathbf{k}|}}\frac{ p^\mu}{p\cdot k}[\mathbf{b}^\dagger_{p} \mathbf{b}_{p-k} \mathbf{a}_{\mu k}(E_p-E_{p-k})  +\mathbf{b}^\dagger_{p-k}\mathbf{b}_{p}\mathbf{a}^\dagger_{\mu k} (E_p-E_{p-k})]  \notag \\ \notag \\
&=- \frac{ie}{(2\pi)^{3/2}} \int \frac{ d^3 k d^3 p}{\sqrt{2 |\mathbf{k}|}}\frac{ p^\mu}{p\cdot k}\frac{\mathbf{p \cdot k}}{E_p}[\mathbf{b}^\dagger_{p} \mathbf{b}_{p-k} \mathbf{a}_{\mu k}  +\mathbf{b}^\dagger_{p-k} \mathbf{b}_{p}\mathbf{a}^\dagger_{\mu k}]  +O(\alpha^{5/2}).
\end{align}
We used $k \sim \alpha m$ to simplify the difference in energies.  The lowest order term in this difference contributes a factor of $\alpha$, making this commutator $O(\alpha^{3/2})$, rather than $O(\alpha^{1/2})$, as we might have expected.  The commutator of S with the $d^\dagger d$ part of the free fermion Hamiltonian is identical up to a negative sign after taking $b\rightarrow d$.  Combining the two parts, we find
\begin{align}
[S, H_{0f}] &=- \frac{ie}{(2\pi)^{3/2}} \int \frac{ d^3 k d^3 p}{\sqrt{2 |\mathbf{k}|}}\frac{ p^\mu}{p\cdot k}\frac{\mathbf{p \cdot k}}{E_p}[\boldsymbol{\rho}^\dagger(p,k) \mathbf{a}_{\mu k}  +\boldsymbol{\rho}(p,k) \mathbf{a}^\dagger_{\mu k}] . 
\end{align}
The commutator of $S$ with the free photon Hamiltonian is straightforward:
\begin{align}
[S,H_{0ph}] &=\frac{ie}{(2 \pi)^{3/2}}\int \frac{ d^3 k d^3 p}{\sqrt{2 |\mathbf{k}|}}|\mathbf{k}|\frac{p^\mu}{p \cdot k}(\mathbf{a}_{\mu k} \boldsymbol{\rho}^\dagger (p,k)+\mathbf{a}^\dagger_{\mu k} \boldsymbol{\rho} (p,k)),
\end{align}
The sum of the two terms multiplied by $-i$ is
\begin{align*}
-i[S,H_{0}]&=-i\left(\frac{ie}{(2 \pi)^{3/2}}\int \frac{ d^3 k d^3 p}{\sqrt{2 |\mathbf{k}|}}\left(\frac{E_p|\mathbf{k}|-\mathbf{p\cdot k}}{E_p}\right) \frac{p^\mu}{p \cdot k}(\mathbf{a}_{\mu k} \boldsymbol{\rho}^\dagger (p,k)+\mathbf{a}^\dagger_{\mu k} \boldsymbol{\rho} (p,k))\right) \\ \\
&=\frac{e}{(2 \pi)^{3/2}}\int \frac{ d^3 k d^3 p}{E_p \sqrt{2 |\mathbf{k}|}}p^\mu(\mathbf{a}_{\mu k} \boldsymbol{\rho}^\dagger (p,k)+\mathbf{a}^\dagger_{\mu k} \boldsymbol{\rho} (p,k)) \\ \\
&=-H_{Ias}.
\end{align*}
which confirms \eqref{eq:18}. 

\flushleft
{\Large{\textbf{Acknowledgements}}}\\
~
We thank Juha \"Ayst\"o for his support at the Helsinki Institute of Physics and Paul Hoyer for instructive and stimulating discussions. This work was supported in part by the Maryland Center for Fundamental Physics.

\end{document}